\documentclass[aip,amsmath,amssymb,eprint]{revtex4-2}
\usepackage{graphicx}
\usepackage{dcolumn}
\usepackage{bm}
\usepackage[utf8]{inputenc}
\usepackage[T1]{fontenc}
\usepackage{xcolor}
\usepackage{mathptmx}
\usepackage{comment}
\usepackage{etoolbox}
\usepackage{float}

\makeatletter
\def\@email#1#2{%
	\endgroup
	\patchcmd{\titleblock@produce}
	{\frontmatter@RRAPformat}
	{\frontmatter@RRAPformat{\produce@RRAP{*#1\href{mailto:#2}{#2}}}\frontmatter@RRAPformat}
	{}{}
}%
\makeatother

\begin{document}
	
	
\title[]{Carrier Dynamics in High-density Photo-doped MoS$_2$: Monolayer vs Multilayer}

\author{Durga Prasad Khatua}
\affiliation{Nano Science Laboratory, Materials Science Section, Raja Ramanna Centre for Advanced 
	Technology, Indore, India - 452013.}
\affiliation{Homi Bhabha National Institute, Training School Complex, Anushakti Nagar, 
	Mumbai, India - 400094.}
\affiliation{Currently at Mechanical and Aerospace Engineering Department, University of California, Los Angeles, Los Angeles, CA 90095, USA.}
\author{Asha Singh}
\affiliation{Nano Science Laboratory, Materials Science Section, Raja Ramanna Centre 
	for Advanced Technology, Indore, India - 452013.}
\author{Sabina Gurung}
\affiliation{Nano Science Laboratory, Materials Science Section, Raja Ramanna Centre for Advanced 
	Technology, Indore, India - 452013.}
\affiliation{Homi Bhabha National Institute, Training School Complex, Anushakti Nagar, 
		Mumbai, India - 400094.}
\affiliation{Currently at Department of Chemistry and York Biomedical Research Institute, University of York, York-YO10 5DD, UK.}
\author{J. Jayabalan}
\email{quantumbalan@gmail.com}
\affiliation{Nano Science Laboratory, Materials Science Section, Raja Ramanna Centre 
	for Advanced Technology, Indore, India - 452013.}
\affiliation{Homi Bhabha National Institute, Training School Complex, Anushakti Nagar, Mumbai, 
		India - 400094.}
\affiliation{Currently at Faculty for physics and CENIDE, University of Duisburg-Essen, 47057 Duisburg, Germany.}
  
\date{\today}

\begin{abstract}
Monolayer and multilayer MoS$_2$ are extremely fascinating materials 
for the use in lasers, compact optical parametric amplifiers, and 
high-power detectors which demands high excitation light-matter 
interaction. Consequently, it is essential to understand the carrier 
dynamics in both the cases at such high excitation densities. In this 
work, we investigate the carrier dynamics of monolayer and multilayer 
MoS$_2$ at photo-doping densities around Mott Density. It is observed 
that, despite the similarity in band structure near K-point and 
formation of A-exciton, a substantial difference in the carrier 
dynamics is observed reflecting the influence of the entire band 
structure. The exciton dissociation, bandgap renormalization, and 
intervalley relaxation play a consequential role in dictating the 
ultrafast transient properties of these samples. The study in this paper provide 
a substantial understanding of fundamental optoelectronic properties 
of the two-dimensional MoS$_2$, paving a way for its potential 
applications in various photonic and optoelectronic domain.  
\end{abstract}

\maketitle

\section{\label{sec: Introduction}INTRODUCTION}
Due to the increasing demand for optoelectronics applications and scalability, 
search for two-dimensional layered materials lead to transition metal 
dichalcogenides. Among them, Molybdenum Disulfide (MoS$_2$) is a highly 
studied material because of its abundance in nature. MoS$_2$ shows a transition 
from indirect to direct bandgap while thinning it down from bulk to single layer
\cite{mak2010atomically, splendiani2010emerging}. In single layer form, it possesses
great physical properties that can be used in a wide range of applications, such 
as optoelectronics, valleytronics, and quantum technologies
\cite{lembke2012breakdown, zheng2014tuning, luo2017opto}.
In monolayer, the quantum confinement and Coulombic interaction of 
the charge carriers become significant leading to a high binding energy excitons of the 
order of $\sim$ 0.5 eV \cite{yuan2017exciton, chernikov2014exciton, 
ugeda2014giant}. In contrast, the exciton binding energy in bulk MoS$_2$ is 
$\sim$45 meV \cite{Borzda2015charge,evans1967exciton}. Thus easy exciton dissociation 
is possible in a bulk compared to that of monolayer. Furthermore, MoS$_2$ possesses
three excitons in visible region, named A, B, and C \cite{mak2010atomically, splendiani2010emerging}.
In addition, because of the high 
contribution of the $d$ orbital electrons of the metal atom, it has high spin-orbit 
coupling at the $K$ point of the Brillouin zone, which is strong enough to eliminate 
the degeneracy in the valence and conduction bands. Moreover, monolayer MoS$_2$ has two 
nonequivalent degenerate valleys at the $K$ points due to lack of inversion symmetry. However, the presence of inversion symmetry in a bulk MoS$_2$ makes the bands near $K$ points equivalent with spin and optical transition properties remaining the same as that of monolayer\cite{mak2010atomically,splendiani2010emerging,Borzda2015charge}. 

The usage of bulk and monolayer MoS$_2$ in optoelectronics and sensor devices requires a 
thorough understanding of the fundamental carrier behaviors in the materials. 
Measurement of carrier relaxation dynamics will provide insight into the transient
optical and electronic properties of these samples. Different groups have studied the 
carrier dynamics in bulk and monolayer MoS$_2$ earlier to understand the carrier behavior in 
the materials with different excitation and probing conditions \cite{cha20151s,
schiettecatte2019ultrafast, berkelbach2013theory, Wang2015ultrafast, wang2015surface,
Lee2020annihilation, wang2017slow, tsai2020ultrafast, sun2014observation, zhang2021defect,
Nie2014ultrafast, Nie2015ultrafast, Chi2020observation, Ceballos2016excitons, Wang2021investigation,
Borzda2015charge, pan2020ultrafast, Yu2019inhomogeneous, Volzer2021Fluence, seo2016ultrafast,
liu2019direct, zhang2014absorption}. 
However, so far the focus was more on the dynamics at low excitation densities.

In this work, we present the results of the carrier dynamics studied in monolayer and 
multilayer MoS$_2$ flake at different photoexcited carrier densities from 0.66 $\times$ 
10$^{14}$ cm$^{-2}$ to 4.16 $\times$ 10$^{14}$ cm$^{-2}$ at A-exciton energy. Atomic
Force Microscope (AFM) measurements were performed to identify and confirm the 
monolayer flake and to estimate number of layers in the selected multilayer 
MoS$_2$. The thickness of the multilayer on which the carrier dynamics measurements are 
reported is 8.9 nm, which corresponds to about 11 layers of MoS$_2$ monolayer. 
As previously reported, with increase in number of layers, a gradual change in the optical property of 
the MoS$_2$ is observed\cite{mak2010atomically, wang2015surface}. However, the properties remains almost unchanged if the number
of layers increased beyond six\cite{mak2010atomically, wang2015surface}.
Thus, the present sample with 11 layers represents a bulk case. In this work, by investigating the transient carrier behavior in both the samples, we present a detail carrier relaxation pathways in both the samples. We find that the Auger and defect-assisted Auger processes plays an crucial role in determining carrier relaxation in monolayer whereas an additional process, intravalley scattering dominates in bulk. In the monolayer, intervalley scattering is not prominent as carriers are excited at the lowest transition energy (A-exciton). Furthermore, the peak heights of the transient signal shows a distinct saturation behavior in multilayer at higher pump fluences, where it is linear in monolayer MoS$_2$. These studies are beneficial for the use of MoS$_2$ in variety of optoelectronic device applications.

\section{\label{sec:Result & Discussion}RESULT AND DISCUSSION}
\begin{figure}[h]
\begin{center}	\centerline{\includegraphics[width=\columnwidth]{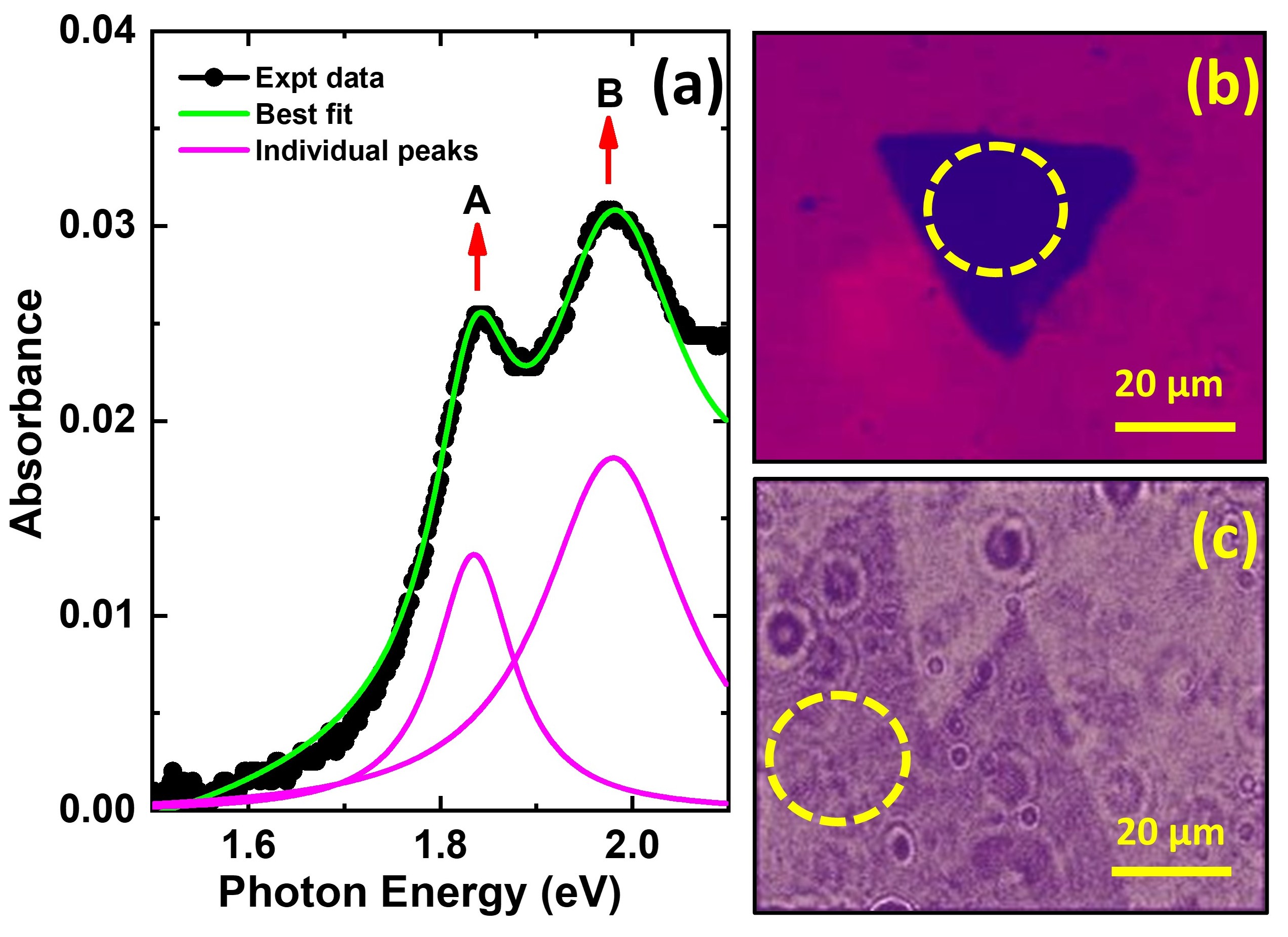}}
	\caption{Optical characterization of the multilayer and monolayer MoS$_2$ flakes. 
		($a$) Averaged absorbance spectrum of the MoS$_2$ flakes. Solid green line 
		represents the best fit to the data using two Lorentzian peaks. Optical image 
		of the ($b$)  multilayer and ($c$) monolayer MoS$_2$ flakes. Dotted circles represent 
		the area where carrier dynamics measurements are performed.}
	\label{Fig01-Sample}
\end{center}
\end{figure}

\begin{figure}[h!]
	\begin{center}
		\centerline{\includegraphics[width=\columnwidth]{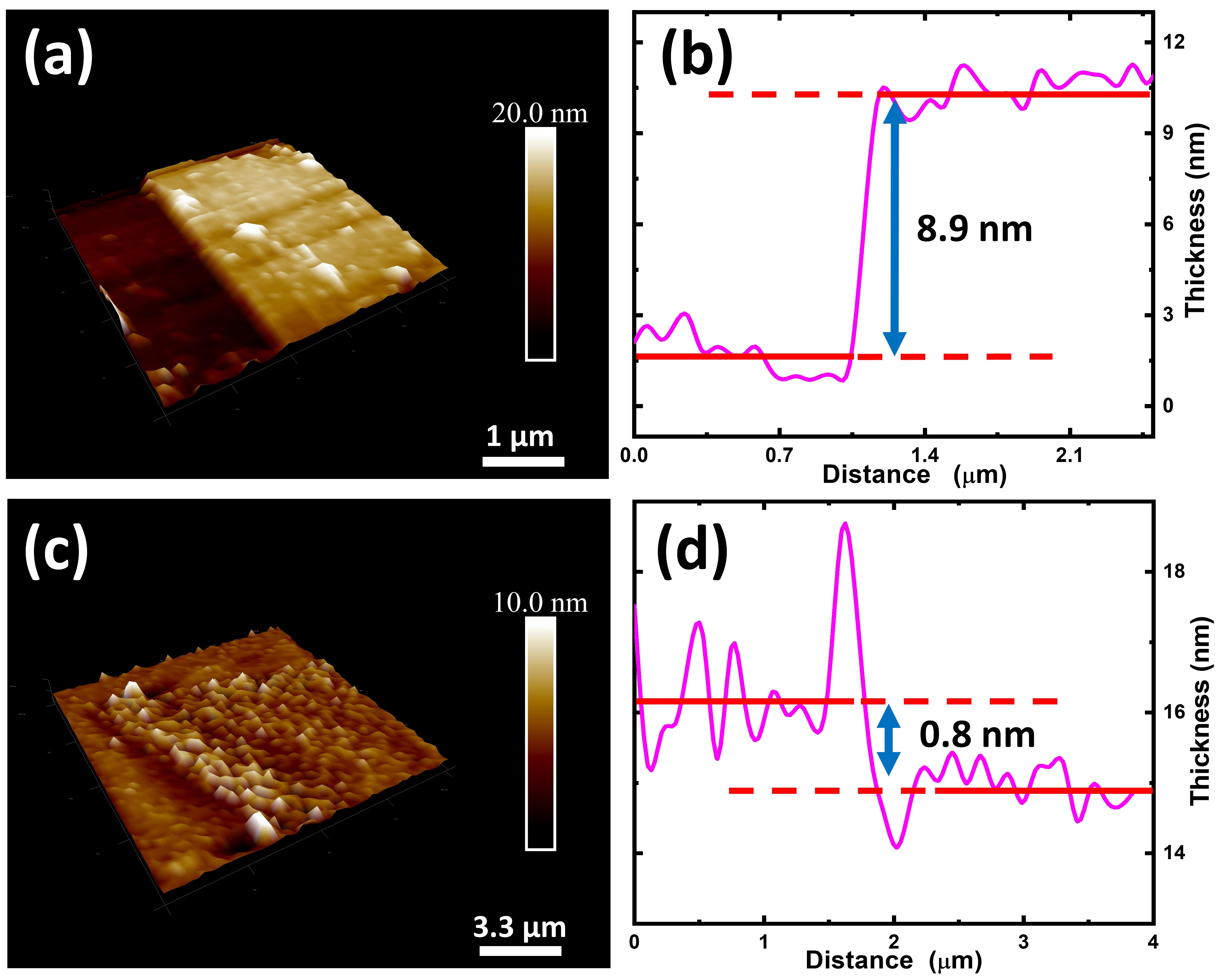}}
		\caption{AFM topographic scan of the ($a$) multilayer and ($c$) monolayer MoS$_2$. 
			AFM line profile of ($b$) multilayer and ($d$) monolayer MoS$_2$.}
		\label{Fig03-AFM}
	\end{center}
\end{figure}

The samples that are used in the ultrafast carrier dynamic measurements are procured 
from 2DLayers, USA, which are grown on sapphire substrate\cite{https://2dlayer.com}. 
Microscopic optical images of the sample show the presence of well-separated and some joint 
MoS$_2$ flakes. Fig.\ref{Fig01-Sample}(a) shows the large area absorption spectrum of 
the flakes which has two distinct peaks in lower energy region corresponding to A and 
B-exciton peaks \cite{splendiani2010emerging, zhang2014absorption,tsai2020ultrafast}. 
These peaks are fitted with Lorentzian functions with peaks at 1.83 eV and 1.98 eV 
and spectral-width of $\sim$ 0.11 eV and $\sim$ 0.18 eV, respectively. The optical images 
of the multilayer and monolayer MoS$_2$ flakes are shown in Figs.\ref{Fig01-Sample}(b) 
and \ref{Fig01-Sample}(c), respectively, which are captured using a microscope 
build with the pump-probe setup \cite{khatua2022excitation, khatua2022comparative}.
The dotted portion in the images shows the area where transient measurements were 
performed. The thickness of the flakes were estimated by performing AFM topography 
scanning measurements at several positions. AFM topography scan of the monolayer and 
multilayer flakes are shown in Figs.\ref{Fig03-AFM}(a) and \ref{Fig03-AFM}(c), 
respectively. Furthermore, the thickness variation of the monolayer and multilayer flakes 
across one edge are shown in Figs.\ref{Fig03-AFM}(b) and \ref{Fig03-AFM}(d), 
respectively. In comparison to the monolayer flake, the multilayer structure 
consists of 11 layers as confirmed from AFM measurements . The in-plane dimension of the multilayer and monolayer flakes are 
$\sim$ 35 $\mu$m and $\sim$ 100 $\mu$m, respectively. 

All the transient measurements were carried out in a standard degenerate pump-probe 
configuration\cite{khatua2020filtering,khatua2022ultrafast}. For these measurements, 
a 35 fs Oscillator-Amplifier-OPA system was used which is operating at 1 kHz repetition 
rate. The full width half maximum diameter of the pump and probe beams at the overlapping 
position are 72 $\mu$m and 24 $\mu$m, which confirms the uniform exposure of the flakes 
to the probe laser beam and thus uniform transient measurements are ensured. The temporal 
width of the laser pulse used is measured to be $\sim$ 60 fs with a spectral with of 
$\sim$ 72 meV at the sample place. In this article, all the measurements are performed 
in the pump fluence range, 1.2 mJcm$^{-2}$ to 7.5 mJcm$^{-2}$ which corresponds to 
carrier density of 6.6 $\times$ 10$^{13}$ cm$^{-2}$ to 4.1 $\times$ 10$^{14}$ cm$^{-2}$ 
at A-exciton wavelength\cite{khatua2022ultrafast, khatua2022excitation, khatua2022comparative}.

Since, all measurements are carried out at such high excitation densities, it is 
essential to ensure that there is no damage in the flakes at such high density.
Pan {\it et al.} carried out this type of experiments on multilayer sample and 
found out that up to 150 mJcm$^{-2}$ pump fluence, the samples remained intact. 
When they increased the fluence up to 400 mJcm$^{-2}$, nanoridges and nanocracks 
were formed in the flakes\cite{pan2020ultrafast}. To check the stability of the 
present samples, we exposed it up to 50 mJcm$^{-2}$ in increasing fluence steps 
of 10 mJcm$^{-2}$ for 10 min. The AFM topography images were recorded after exposure
to each fluence. From these measurements, we found that up to 50 mJcm$^{-2}$ there was 
no structural changes observed in the samples\cite{khatua2022ultrafast, 
khatua2022comparative}. In the current study, the maximum pump fluence used is 
7.5 mJcm$^{-2}$ which is almost 13 times less than the damage threshold.

\begin{figure*}[!ht]
\center
\includegraphics[width=\linewidth]{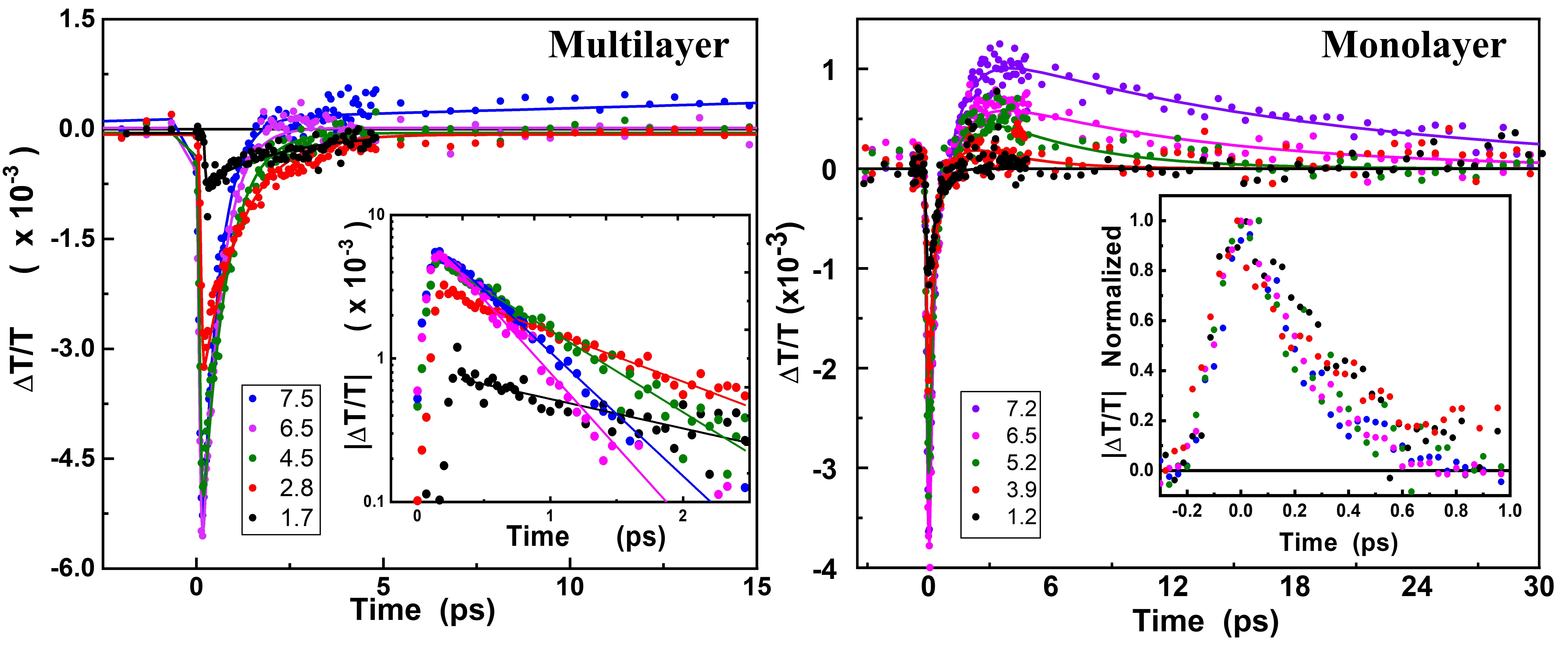}
\caption{Pump fluence dependent transient transmission signal recorded by pumping and 
	probing at 672 nm: ($a$) Multilayer MoS$_2$ flake (Inset: same data in short time 
	scale and with y-axis in natural log scale) and ($b$) monolayer MoS$_2$ flake (Inset: 
	normalized data in short time scale.). The legend shows the pump fluence in 
	mJcm$^{-2}$.}
\label{Fig04-TransientBulkML}
\end{figure*}

Fluence dependent transient transmission signal ($\Delta T/T$) of the multilayer MoS$_2$ 
flake is shown in Fig.\ref{Fig04-TransientBulkML}(a). The transmission of the 
sample reduces with the arrival of the pump pulse and attains a negative peak 
nearly by the end of the pump pulse. This change in the transmission then starts 
recovering as the delay between the pump and probe increases and returns to its 
original unperturbed transmission state by about 5 ps. As the pump fluence increases, 
the magnitude of the peak change in the transmission also increases. At higher fluence, 
the recovery of transmission was found to be faster than that at lower pump fluences. 
Furthermore, when pump fluence is $<$ 6.5 mJcm$^{-2}$, the transient signal recovers 
completely to zero level by about 5 ps. However, as the fluence increases beyond 6.5 
mJcm$^{-2}$, $\Delta T/T$ was found to change its sign from negative to positive after 
4 ps. This positive change in transient transmission stays for about tens of picoseconds. 
In the inset of Fig.\ref{Fig04-TransientBulkML}(a), we show the same transient signal 
of the multilayer MoS$_2$ flake for the first few picoseconds time scale, while y-axis is 
in log scale. The solid lines in the plot are the best linear fit to the data in the 
0.3 ps to 2.5 ps range. In Fig.\ref{Fig04-TransientBulkML}(b), the transient transmission 
signal of a monolayer MoS$_2$ recorded in the A-exciton state is shown for the same pump fluence 
range as that of the multilayer. As in the multilayer, the $\Delta T/T$ 
signal of the monolayer MoS$_2$ starts 
decreasing with the arrival of the pump pulse showing a negative peak within the laser 
pulse width. This peak starts recovering with increasing delay time. As the fluence 
increases, the negative peak height of the $\Delta T/T$ increases. For pump fluences lower
than 4.1 mJcm$^{-2}$, the transient signal recovers completely within first 2 ps. 
However, at higher pump fluences, $\Delta T/T$ changes sign from negative to positive. 
This positive signal builds up slowly and attained a peak by about 3 ps which is also 
shorter when compared to that of the multilayer. In the inset of Fig.\ref{Fig04-TransientBulkML}(b), 
modulus of the same $\Delta T/T$ signal of monolayer MoS$_2$ is shown in short time scale 
with normalization. From this plot, it can be seen that the relaxation time of the 
initial negative transient signal decreases at higher pump fluences. Further, it should 
be noted that positive change in $\Delta T/T$ at later time appears earlier in monolayer MoS$_2$ 
compared to multilayer.

\begin{figure}[h]
	\begin{center}
		\centerline{\includegraphics[width=\columnwidth]{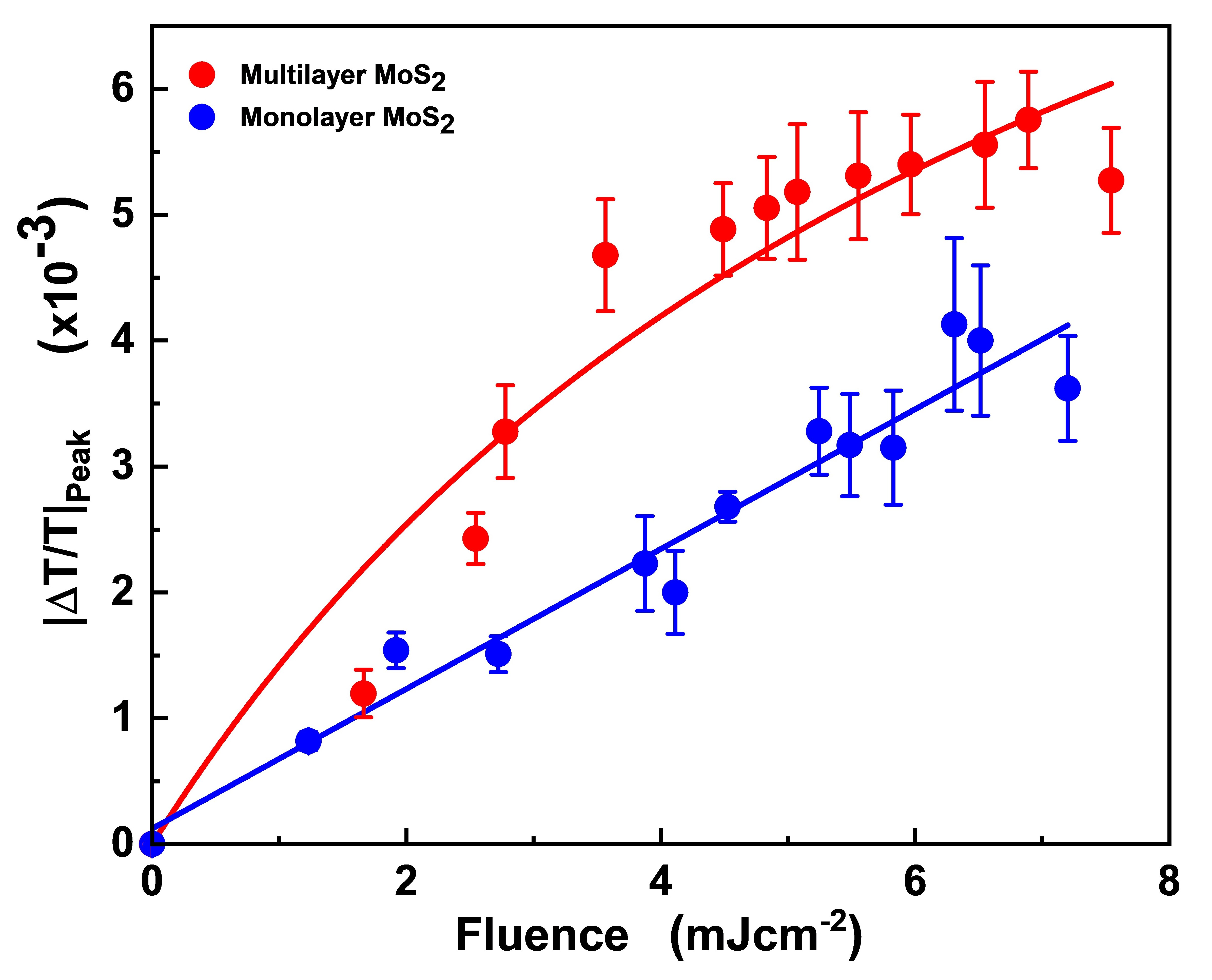}}
		\caption{Pump fluence dependence of the magnitude initial negative peak of 
			$\Delta T/T$ for the multilayer MoS$_2$ (red dots) and for the monolayer MoS$_2$ 
			(blue dots). Solid lines represents the best fit to the corresponding data 
			sets.}
		\label{Fig05-PeaksBulkML}
	\end{center}
\end{figure}

Fig.\ref{Fig05-PeaksBulkML} shows the variation of the magnitude of the initial 
negative peak height of $\Delta T/T$ ($|\Delta T/T|_{Peak}$) with pump fluence for 
the multilayer MoS$_2$ (red dots) and monolayer MoS$_2$ (blue dots). In case of multilayer MoS$_2$, 
initially, the peak height increases rapidly with pump fluence up to 4 mJcm$^{-2}$. 
As pump fluence increases beyond 4 mJcm$^{-2}$, the transient signal starts showing 
saturation. In contrast, variation of the initial peak height of $\Delta T/T$ in 
monolayer MoS$_2$ with pump fluence does not show any saturation in the range of pump 
fluence used in the measurement. A linear dependence of the peak change of 
$\Delta T/T$ with pump fluence is observed in monolayer MoS$_2$ while it is 
non-linear in multilayer. The solid blue line represents the linear fit to the 
$|\Delta T/T|_{Peak}$ data of monolayer MoS$_2$. In multilayer case, the data is fitted with a 
saturation function,
\begin{equation}\label{Eq:SaturationFitBulkData}
	\mathcal{T}(F) = \dfrac{PF}{QF+R} , 
\end{equation}
where $\mathcal{T}$ is the magnitude of the peak change in the transient signal, 
$F$ is the pump fluence, and $P$, $Q$, and $R$ are constants.

\begin{figure}[h]
	\begin{center}
		\centerline{\includegraphics[width=\columnwidth]{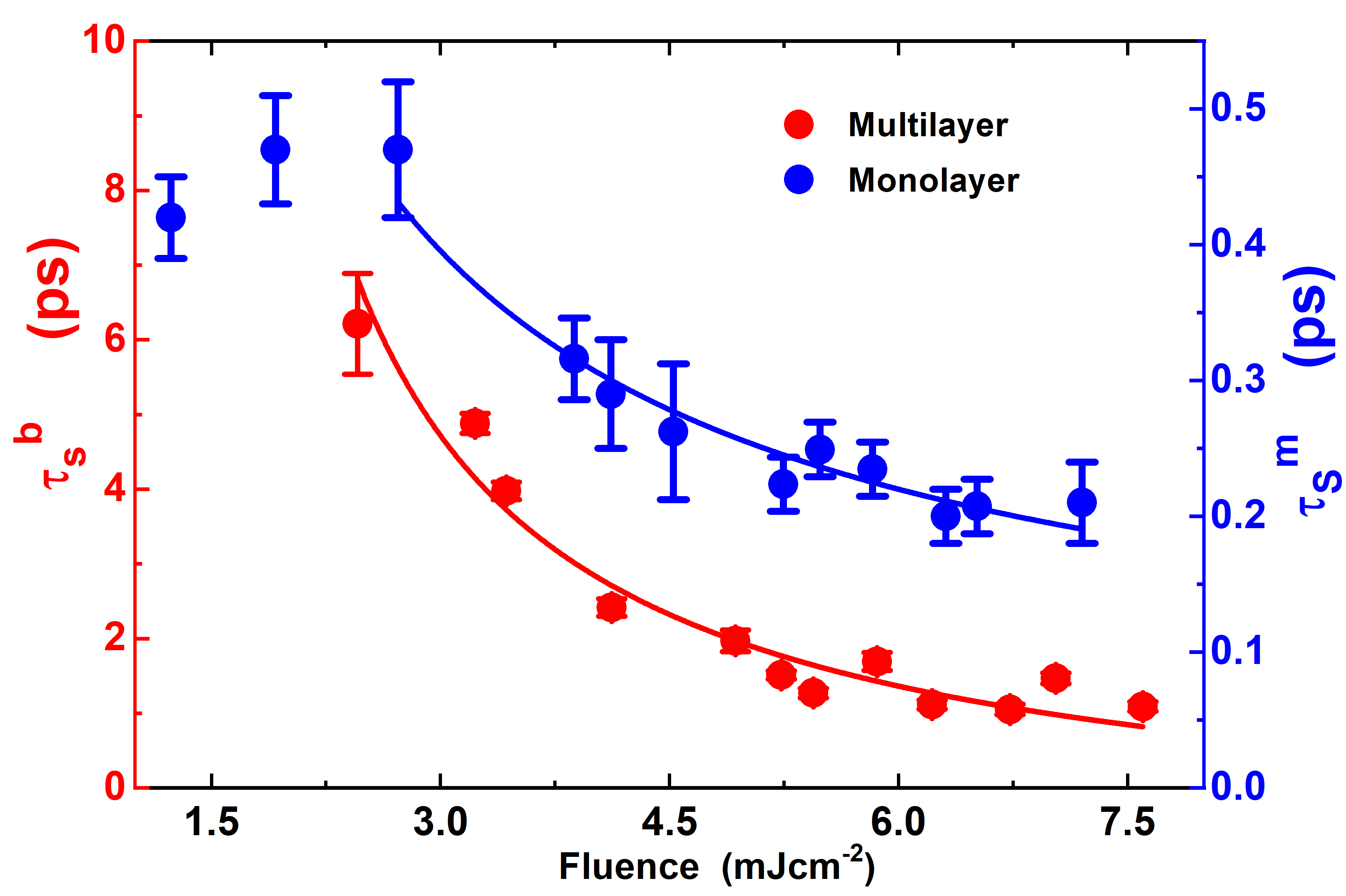}}
		\caption{Pump fluence dependent relaxation time of the multilayer (red dots) and 
			monolayer MoS$_2$ (blue dots) flakes. The solid lines are the corresponding best
			fits to the data.}
		\label{Fig06-DecayTimeBulkML}
	\end{center}
\end{figure}

The relaxation time of the initial peak for the multilayer case ($\tau_s^b$) is 
extracted by fitting the transient data in the range of 0.3 ps to 2.5 ps to a 
single exponential function for each fluence. The obtained best fit decay times 
for various fluence is shown in Fig.\ref{Fig06-DecayTimeBulkML}. We find that the
fluence dependence of $\tau_s^b$ can be fitted with a function $\tau = A/F + B$, 
where, $A$ and $B$ are constants. The obtained best fit is also shown in 
Fig.\ref{Fig06-DecayTimeBulkML}.

The time dependence of the monolayer case is discussed in good details in our 
earlier publications\cite{khatua2022excitation, khatua2022comparative}. However,
for the sake of completeness, in short we discuss the time dependence of 
monolayer sample here. To obtain the decay time of $\Delta T/T$ for the 
monolayer and multilayer MoS$_2$ case, we fit the experimental data to an empirical function, 
\begin{eqnarray}
	\mathcal{F} &=& A_0 \left[ 1 + \mathcal{E} \left( \frac{t}{t_0} \right)
	\right] \exp{\left( -\frac{t}{\tau^{m}_s} \right)} \nonumber\\
	&\quad& \quad + B_0 \left[ 1 - \exp{ \left( \frac{t-t_d}{\tau_r} \right) } 
	\right] \exp{ \left(-\frac{t-t_d}{\tau_L} \right) }.
	\label{Eq:FitALBulkML}
\end{eqnarray}
The first part of the equation represents the initial fast change in the transmission
due to the pump intensity and a corresponding exponential decay. The second part in the 
equation is for fitting the slow exponential increase in the transmission and the 
corresponding decay. For comparison with that of multilayer MoS$_2$, the initial 
relaxation time of monolayer MoS$_2$ ($\tau^{m}_s$) is also plotted in 
Fig.\ref{Fig06-DecayTimeBulkML}. At lower pump fluences ($<$ 2.7 mJcm$^{-2}$),
$\tau^{m}_s$ remains almost constant around 0.4 ps. However, as the pump fluence 
is increased beyond 2.7 mJcm$^{-2}$, the relaxation time decreases rapidly showing
a similar trend as that of the multilayer case.

\begin{figure}[h]
	\begin{center}
		\centerline{\includegraphics[width=\columnwidth]{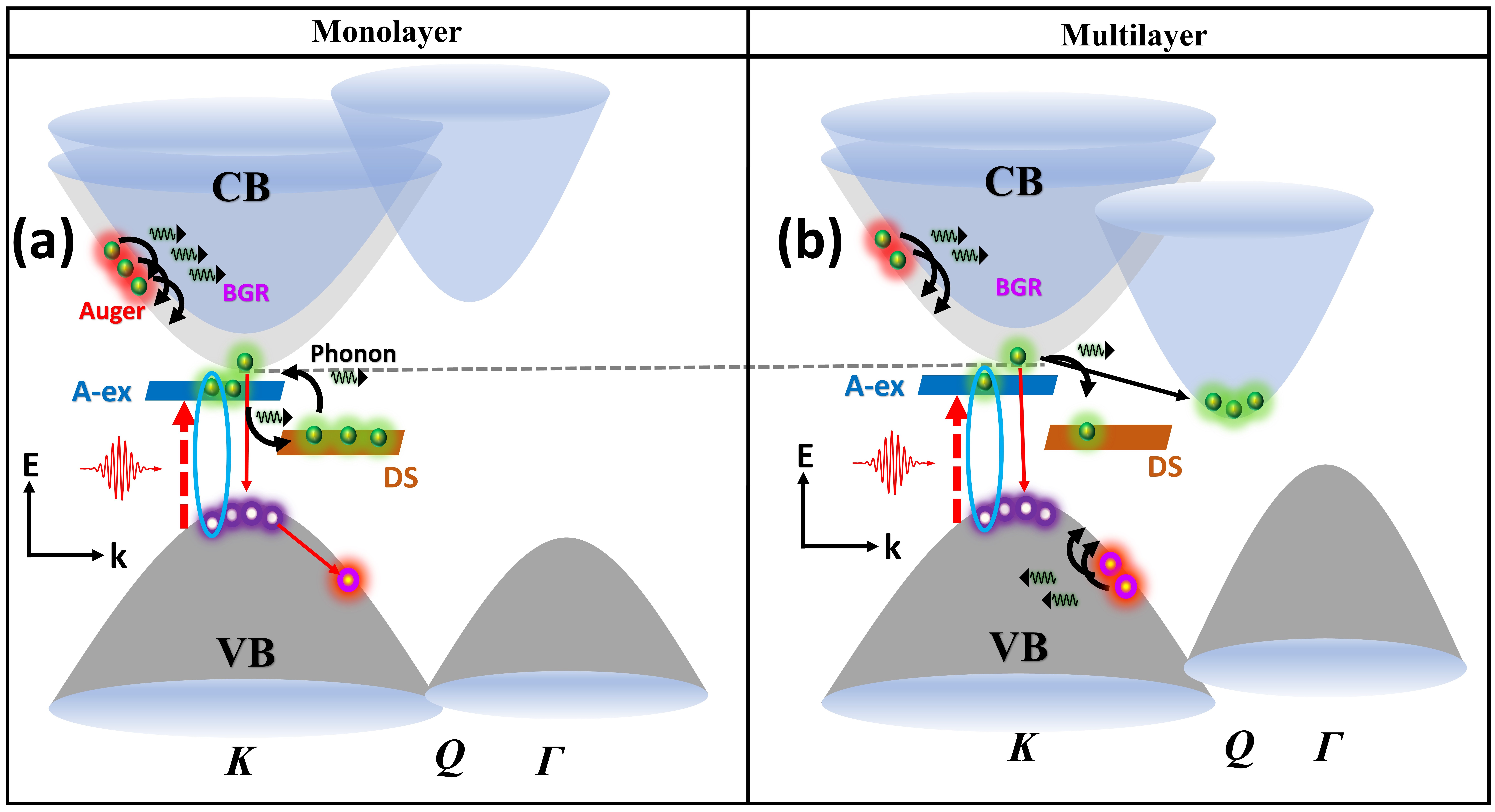}}
		\caption{Schematic of the proposed processes: ($a$) Processes in a 
		monolayer MoS$_2$ sample while exciting the carriers at A-exciton leading 
		to exciton dissociation, Auger scattering, and intervalley scattering. 
		($b$) Processes in a bulk MoS$_2$ near A-exciton transition leading to 
		exciton dissociation, Auger scattering.}
		\label{Fig07-Process}
	\end{center}
\end{figure}

It is well known that the electronic band structure of multilayer MoS$_2$ is very different 
form that of monolayer MoS$_2$. Fig.\ref{Fig07-Process} shows the schematic of 
the band structure of a multilayer and monolayer case. The bandgap of a monolayer is 
1.83 eV and happens to be at the $K$ valley \cite{mak2010atomically, splendiani2010emerging}. 
If there are two layers of MoS$_2$, the conduction band in between $\Gamma$ 
and $K$ valley ($Q$ point) starts reducing, reaching an energy value which is lower 
than that of at $K$ point \cite{mak2010atomically}. Such lowering of energy makes the 
two-layer MoS$_2$ itself an indirect bandgap material. The photoluminescence efficiency 
of double-layer MoS$_2$ thus gets reduced by about a factor of $\sim$ 40 when compared 
to that of a monolayer MoS$_2$ \cite{mak2010atomically}. With increasing layer number, 
the conduction band energy in between $\Gamma$ and $K$ point reduces even more
\cite{mak2010atomically,wang2015surface}. After 5 to 6 layers, the changes in bandgap 
stabilizes resulting in a band structure which is independent of the number of layers. 
Thus the multilayer flake studied here is an indirect bandgap semiconductor flake. 
Despite their differences, the excitonics resonance energy of bulk MoS$_2$ remains 
nearly the same at the K point as that of monolayer MoS$_2$ at 1.83 eV 
\cite{Borzda2015charge,mak2010atomically}. Further, in case of a bulk sample, the 
binding energy of A-exciton is around $\sim$ 45 meV making exciton dissociation in bulk 
easier compared to a monolayer\cite{Borzda2015charge, evans1967exciton}.

Now let us look at the physical processes that drive the transient changes in optical 
response of the monolayer and bulk flakes after excitation by a short-pulse. 
Different groups have earlier studied carrier dynamics in bulk and monolayer MoS$_2$ flakes at 
the A-exciton transition energy and reported various physical phenomena that dominate for
few hundred of picoseconds after an ultrafast pulse excitation. 
The ultrafast optical response of the bulk and monolayer MoS$_2$ flakes 
are also expected to be different from each other due to the difference 
in electronic band structure leading to dominance of different processes.
The observed physical phenomena are highly dependent on the excitation photon 
energy, probing photon energy as well as the type of samples, excitation density, and delay time. 
Various processes like exciton formation\cite{Chi2020observation, Yu2019inhomogeneous, 
Ceballos2016excitons}, exciton dissociation\cite{Nie2014ultrafast, Nie2015ultrafast,
khatua2022ultrafast,khatua2022excitation}, exciton-exciton annihilation\cite{Lee2020annihilation,
wang2017slow, tsai2020ultrafast,sun2014observation,zhang2021defect}, carrier capture 
to defect and trap states \cite{cha20151s,schiettecatte2019ultrafast,seo2016ultrafast}, 
Auger and defect-assisted Auger scattering \cite{zhang2014absorption, berkelbach2013theory, Wang2015ultrafast,wang2015surface}, hot-phonon reabsorption \cite{Wang2021investigation, 
He2020cross}, and intervalley scattering \cite{wallauer2016intervalley,kumar2013charge} 
are likely to happen in both bulk and monolayer MoS$_2$ flakes. These processes will lead 
to band bleaching \cite{Chernikov2015population}, band gap renormalization (BGR)
\cite{He2020cross,pogna2016photo}, exciton peak shift and broadening\cite{Wang2021investigation,
wang2017slow, sim2013exciton}, etc. Due to these phenomena, the optical properties of 
the material will change leading to a change in the transmission of the probe beam. 

In the present study, carriers are excited directly at the A-exciton transition energy. 
However, at A-exciton transition energy, the absorption coefficient of a monolayer MoS$_2$ is higher than that of bulk 
MoS$_2$ by 40\%, \cite{castellanos2016spatially}. Hence, the excited number density is 
higher in monolayer sample than that of bulk MoS$_2$ by about 40\%.
Due to higher excited carrier density, the monolayer is expected to show higher 
$|\Delta T/T|_{peak}$ which is contradicting to the observed $|\Delta T/T|_{peak}$ (Fig. 4). However, the exciton 
binding energy in bulk is much lower than that of monolayer, such that exciton dissociation 
can occur at a higher rate in bulk when compared to that of monolayer \cite{Nie2014ultrafast, 
Nie2015ultrafast,Borzda2015charge}. Such exciton dissociation create more free carriers in bulk 
flake, despite having lower absorption coefficient than monolayer. 
Thus, a larger BGR is expected to occurs in bulk compared to that of monolayer MoS$_2$ 
flake for the same excitation density. Therefore, we observed higher $|\Delta T/T|_{Peak}$ in 
bulk MoS$_2$, when compared to that of monolayer, as shown in Fig.\ref{Fig05-PeaksBulkML}.
Thus the higher $|\Delta T/T|_{Peak}$ observed in multilayer sample is attributed to 
the higher dissociation ability of excitons in bulk. Similar to the case of monolayer, 
over time the free carriers in bulk are expected to decay to defect states and also form 
hot carriers via Auger and defect-assisted Auger processes\cite{cha20151s, 
schiettecatte2019ultrafast,seo2016ultrafast, wang2015surface}. In addition to that, a large 
number of electrons decay to $Q$ valley, leaving holes at $K$ valley 
in tens of femtosecond time scale \cite{wallauer2016intervalley}. The life time of electrons 
transported to $Q$ valley is expected to be the order of several 
picoseconds\cite{wallauer2016intervalley, kumar2013charge}. Thus, even though there are 
relaxation to defect states, a large number of carriers will get retained in the conduction 
band as free carriers in bulk. Further, since carriers are also now at different valleys, 
the Auger and defect-assisted Auger are less efficient in case of the multilayer sample. 
In addition, in bulk MoS$_2$, due to its higher thickness compared to monolayer, screening 
of the dielectric is stronger and quantum confinement effect is weaker. This leads to decrease 
in strength of Coulomb interaction which in turn leads to weaker Auger 
scattering process \cite{bera2021atomlike,sim2021opposite}.
Hence, the recovery of the transient change in the transmission is expected to be much longer 
in bulk when compared to that of monolayer. This explains why the recovery in bulk takes much 
longer time as observed in the experiment (Fig.\ref{Fig06-DecayTimeBulkML}). 
For example, at 2.5 mJcm$^{-2}$, the decay time of bulk is $\sim$ 4.8 ps while in the case 
of monolayer, it is 0.4 ps. Similarly, at about 7 mJcm$^{-2}$ pump fluence, decay time 
of bulk is about 1 ps while for the case of monolayer, it is 0.2 ps. Further, intervalley 
scattering and relaxation of carriers from there reduce the possibility of formation of 
A-excitons. This is the reason why the bleaching of $\Delta T/T$ in bulk MoS$_2$ case is 
much weaker and observed only at much larger excitation fluence.
  
\section{Conclusion}
In this article, we showed that the ultrafast response of monolayer and bulk MoS$_2$ 
are different when excited by ultrashort pulse. These studies were carried out at 
photoexcitation densities in the range of $\sim$ 6.6 $\times$ 10$^{13}$ cm$^{-2}$ 
to $\sim$ 4.16 $\times$ 10$^{14}$ cm$^{-2}$. In spite of the fact that the bulk 
has a lower excitation density as compared to monolayer, it still showed a higher 
change in $|\Delta T/T|_{Peak}$ when compared to that of monolayer MoS$_2$, since, the 
dissociation of excitons is much easier in bulk due to the lower exciton binding energy. 
Both bulk and monolayer show Auger and defect-assisted Auger recombination and a positive 
change in $\Delta T/T$ at later times, but the bulk shows a much longer recovery time 
and a very low positive change in $\Delta T/T$. Both of these changes are attributed to the 
carriers scattered to other valleys in the conduction band which increases the carrier 
lifetime by reducing exciton formation and reducing the Auger scattering process. 
All the studies presented here will be helpful for the future use of monolayer and 
multilayer MoS$_2$ in lasers, OPAs, and detectors.

\begin{acknowledgments}
All authors are thankful to Mr. Vijay Singh Dawer for his help during the measurements. 
D.P.K. and S.G. are thankful to RRCAT, Indore for providing financial support under the HBNI PhD 
programme. Author J.J. gratefully acknowledged the funding by the Deutsche 
Forschungsgemeinschaft (DFG, German Research Foundation) through Project 
No. BO1823/12 - FOR 5249 (QUAST).
\end{acknowledgments}
	

\nocite{*}
\bibliography{References}
	
\end{document}